\documentclass[12pt]{article}
\usepackage{graphicx}
\usepackage{longtable}

\input{tcilatex}

\begin{document}

\textbf{Phase transitions, memory and frustration in a}

\textbf{Sznajd-like model with synchronous updating}

\bigskip

Lorenzo Sabatelli$^{1}$$^{,2}$ and Peter Richmond$^{1}$

\bigskip

$^{1}$ Department of Physics, Trinity College Dublin, Ireland

$^{2}$ HIM, IFSC, Dublin, Ireland

\bigskip

E-mail: .\underline {sabatell@tcd.ie}

\bigskip

We introduce a consensus model inspired by the Sznajd Model. The updating is
synchronous and memory plays a decisive role in making possible the reaching
of total consensus. We study the transition between the state with
no-consensus to the state with total consensus.

\bigskip


\bigskip

\textbf{Basic features of Sznajd Model}

It is well understood, within human societies, that it is generally easier 
to change someone's opinion by acting within groups than by acting alone. 
For example, a single person stopping on the street and staring at the sky 
is usually ignored (or perhaps considered eccentric). However, if several 
people stare into the sky, they readily induce others to do the same.

\bigskip

Equally it is not unusual to come across door-to-door sale agents working in 
couples rather than on an individual basis. In democratic electoral 
campaigns, small groups (typically couples) of political party activists may 
visit, door-to-door, potential electors and seek to gain their votes. Trade 
union movements often try to coordinate their actions in order to strengthen 
their position against management than if everybody try to negotiate it 
alone.

\bigskip

The underlying principle, in all of the above-mentioned examples may be 
captured by the famous Abraham Lincoln's injunction: "United we stand, 
divided we fall''. This principle was developed into a computational model 
by Katarzyna Sznajd and her father[1].

The simplest (non-trivial) version of their model can be implemented on a 
two-dimension lattice of spins. Each site carries a spin, $S$ , that is 
either up or down. This represents one of the two possible opinions on any 
question. Two neighbouring parallel spins, i.e. two neighbouring people 
sharing the same opinion, convince their neighbours of this opinion. If they 
do not have the same opinion, then they do not influence their neighbours.

\bigskip

The system evolves from one time step to another through a random sequential 
updating mechanism.

\bigskip

The system always reaches an overall consensus, within a sufficiently long 
time and ends up with all spins up (or down) if the initial fraction of up 
opinions is larger (smaller) than $\raise.5ex\hbox{$\scriptstyle 
1$}\kern-.1em/ \kern-.15em\lower.25ex\hbox{$\scriptstyle 2$} $ [2]. Furthermore, the smaller the size of the lattice, the 
smoother is the transition.

\bigskip

 If the random sequential updating is replaced by a synchronous updating mechanism the possibility of reaching a consensus is reduced quite dramatically.

\bigskip

The updating is performed by going systematically through the lattice to 
find the first member of the pair, then choosing randomly the second member 
of the pair within the neighborhood of the first. Having in this way 
completed the assembly of couples, each agent then orients her/himself 
according to her/his neighbours at time step $t$ . Like-minded couples 
will induce their neighbours to turn to the same state (opinion). However a 
single agent may often belong, simultaneously, to the neighbourhood of more 
than one couple (of likeminded agents). In this case, if the couples have 
different opinions, she/he doesn't know what to do (frustration) and ends up 
doing nothing, i.e. sticks with her/his previous opinion. Frustration may 
prevent the system from reaching total consensus.

\bigskip

\textbf{Memory} 

A feature of the agents in the models discussed above is the complete absence of memory. The past plays no role.

\bigskip

In this note we assume that agents are endowed with memory. For the sake of simplicity, they are all thought to have the same memory span $T$ and updating mechanism is synchronous. Agents are keen to change opinion when in the neighbourhood of a like-minded couple, as in the models above. An agent resorts to her/his individual history when frustration occurs. In that case, the new state turns out to be the most frequent of her/his own $T+1$ most recent ($T$ accounts for the past, one for the present). Of course, this rule proves to be more efficient when $T$ is an even number.

\textbf{Phase transitions}
 
 In a two-dimensional Sznajd model with 
asynchronous updating, the system always ends up at a fixed point: either 
all spins point up or they all point down. When synchronous updating is 
chosen things are different. The system converges to a fixed point (all 
spins up or all spins down), only if the asymmetry  $\Delta p_{}$
 (absolute difference) in the initial distribution of opinions is above a 
critical value  $\Delta p_{c}$ that depends not only on the size of the 
lattice [3] but also on the memory length $T$ [fig.1].

\bigskip

 Fig.2 shows 1-$\Delta p_{c}$  as a function of $L$ (the lattice side size). The slope decreases as $T$ increases. For small values of $L$ 
the three curves displayed seem to converge to the same point. This is 
easily understood because for small lattices one expects the impact of 
frustration on total consensus to be small. As a consequence, the role of 
the memory $T$ is not as important as it would be for large values of $L$ .

For large $L$ the curves monotonically decay to 
zero, this suggests that asymptotically (for infinitely large lattice) no 
consensus is possible.

\bigskip

Fig.3 shows  $\Delta p_{c}$  as a function of $T$ . The slope of the two curves appears to be independent of $L$ - an aspect that requires further 
study.

\bigskip

\textbf{Conclusions}

We have studied a model based on the Sznajd consensus model with synchronous 
updating that includes memory. This feature plays a very important role and 
helps overcome frustration and achieve total consensus.

\bigskip

The memory length, $T$ , affects the phase transition from the 
no-consensus state to the total consensus state. The bigger $T$ the 
closer to zero is the critical point. For a given side size $L$ and $T$ enough large an always-consensus situation, as with the random 
sequentially updated Sznajd model, may be attained. Anyway, for infinitely 
large (but with finite memory) lattices, no consensus is possible, just as 
in the case without memory.

\bigskip

\bigskip

\textbf{Acknowledgments}

We would like to thank Dietrich Stauffer who provided continual advice, 
stimulating remarks and good humour during this study.

\bigskip

The authors acknowledge support from the EU via Marie Curie Industrial
Fellowship MCFH-1999-00026

\bigskip

\bigskip

\textbf{References}

[1] K. Sznajd-Weron and J. Sznajd. Opinion Evolution in Closed Community.
Int. J. Mod. Phys. C 11: 1157-1166 (2000).

\bigskip

[2] D. Stauffer. Monte Carlo Simulations of the Sznajd model, Journal of
Artificial Societies and Social Simulation 5, No.1, paper 4 (2002)
(jasss.soc.surrey.ac.uk).

\bigskip

[3] D. Stauffer. Frustration from Simultaneous Updating in Sznajd Consensus
Model. cond-mat/0207598

\newpage

\bigskip

\par Fig. 1.1
\par The phase transition from a no-consensus state to a total consensus state is driven by the parameter $\Delta $p for
 $L$ =17 and $T$ =0, 2, 8. The transition point is shifted towards zero as the agent memory length \$ T\$ increases.

\bigskip

\par Fig. 1.2
\par The phase transition from a no-consensus state to a total consensus state is driven by the parameter $\Delta $p for $L$=101 and $T$=0, 2, 8. The transition point is shifted towards zero as the agent memory length $T$ increases.

\bigskip

 \par Fig. 1.3
\par The phase transition from a no-consensus state to a total consensus state is driven by the parameter $\Delta $p for $L$=301 and
 $T$=0, 2, 8 The transition point is shifted towards zero as the agent memory length $T$ increases.

\bigskip

\par Fig. 1.4
\par The phase transition from a no-consensus state to a total consensus state is driven by the parameter $\Delta $p for $L$=1001 and
 $T$=0, 2, 8. The transition point is shifted towards zero as the agent memory length $T$ increases.

\bigskip

\par Fig. 1.5 
\par The phase transition from a no-consensus state to a total consensus state is driven by the parameter $\Delta $p for $L$=50 and
 $T$=0, 2, 8, 100, 500, 1000. The transition point is shifted towards zero as the agent memory length $T$ increases.

\bigskip

\par Fig. 2
\par Variation with $L$ of 1-$\Delta $p (one minus the absolute value of the difference in the initial probabilities for +1 and --1) for which in half of the cases a consensus was reached. That may be seen as the phase transition point from the state without consensus to the state with consensus. The estimated slope is --0.39 for $T$=0, -0.21 for $T$=2, - 0.11 for $T$=8.

\bigskip

\par Fig. 3
\par Variation with $T$ of the difference in the initial probabilities for which in half of the cases a consensus was reached. That may be seen as the phase transition point from  the state without consensus to the state with consensus. The estimated slope is 0.46 for $L$=17 and 0.47 for $L$=50.

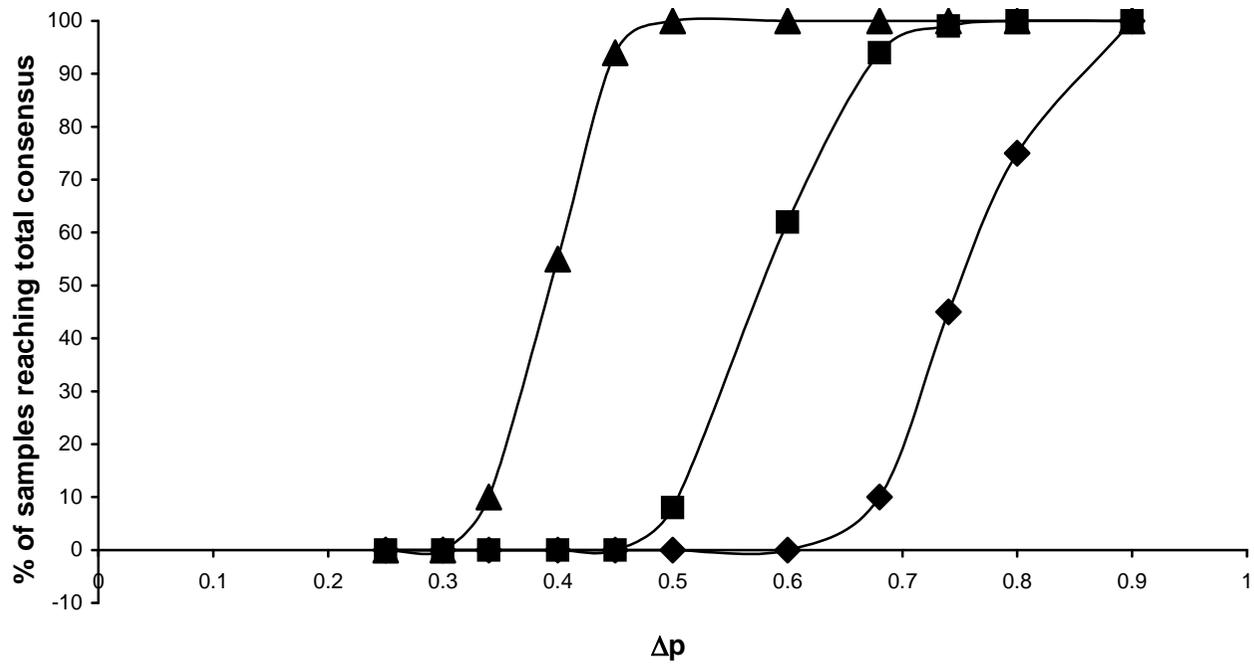
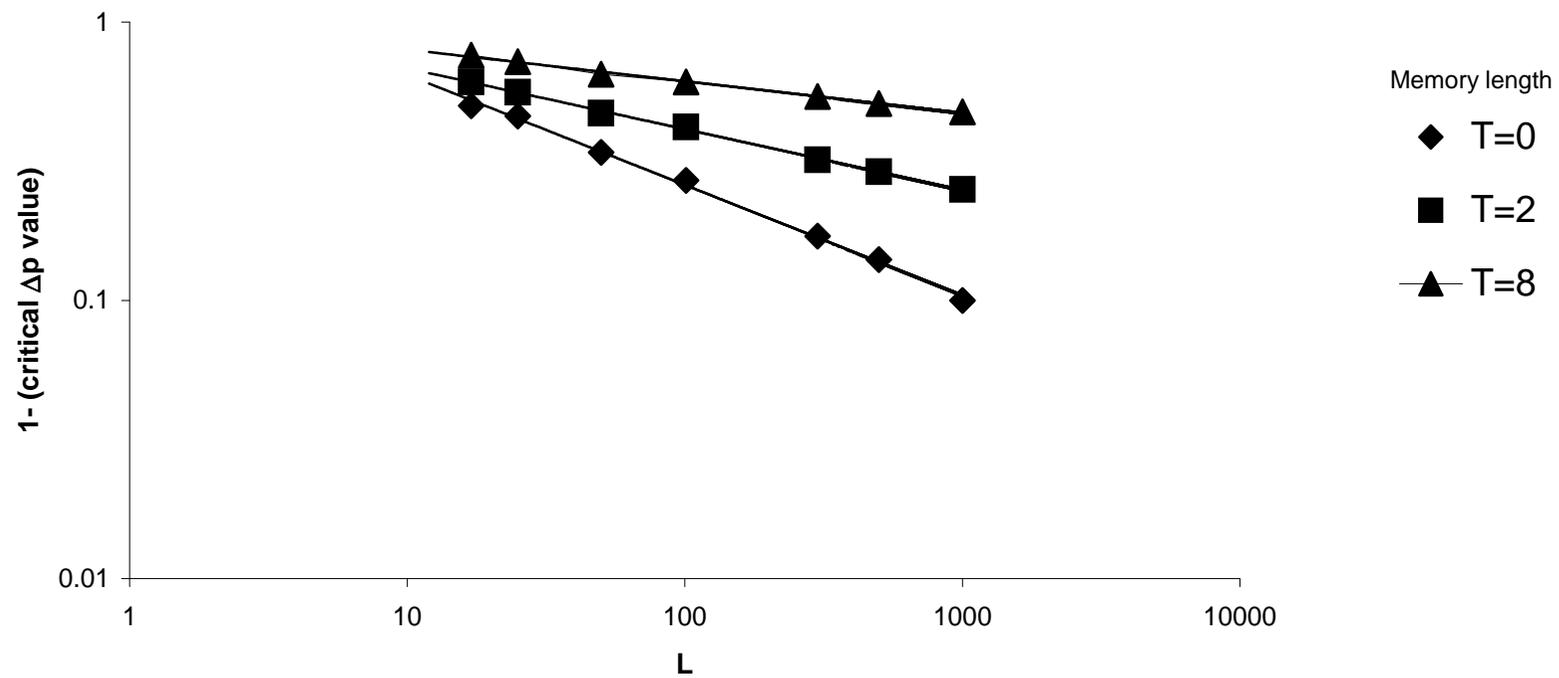
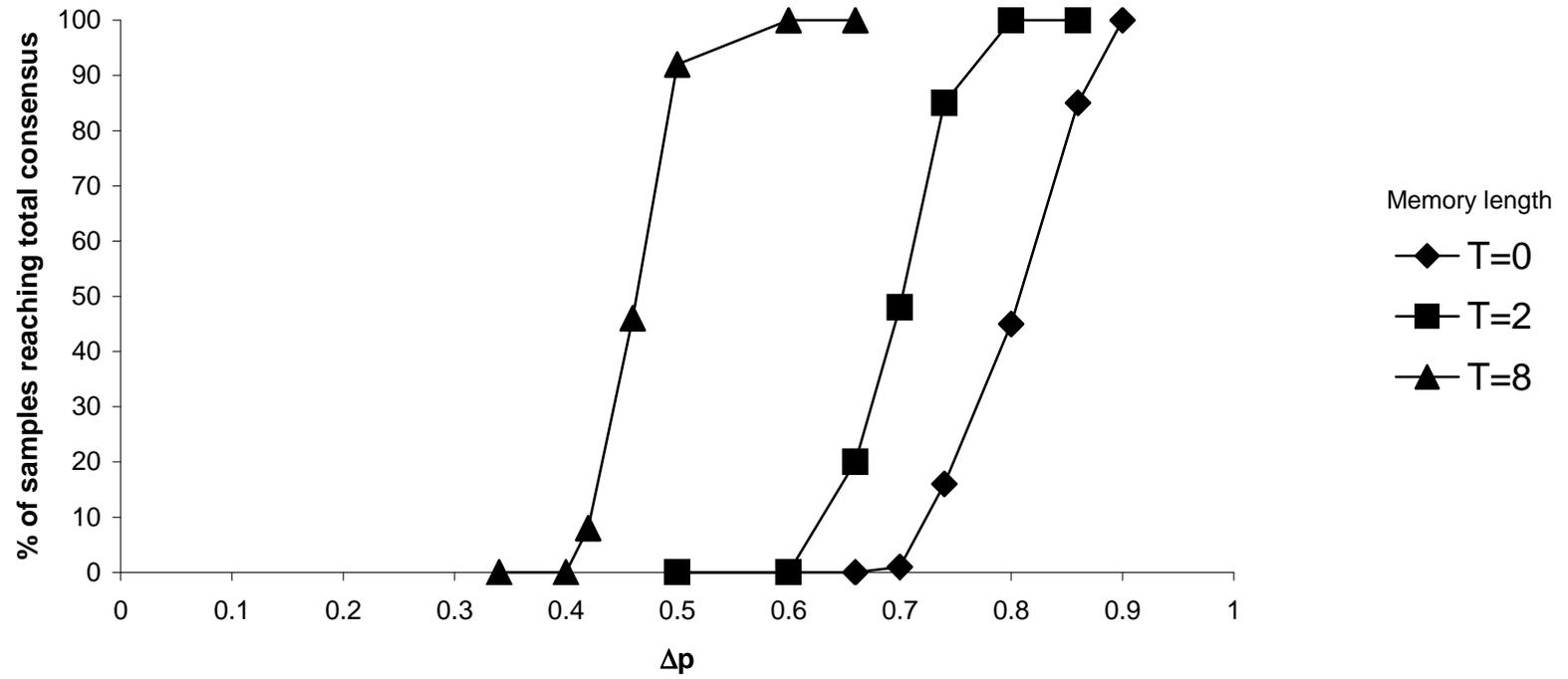
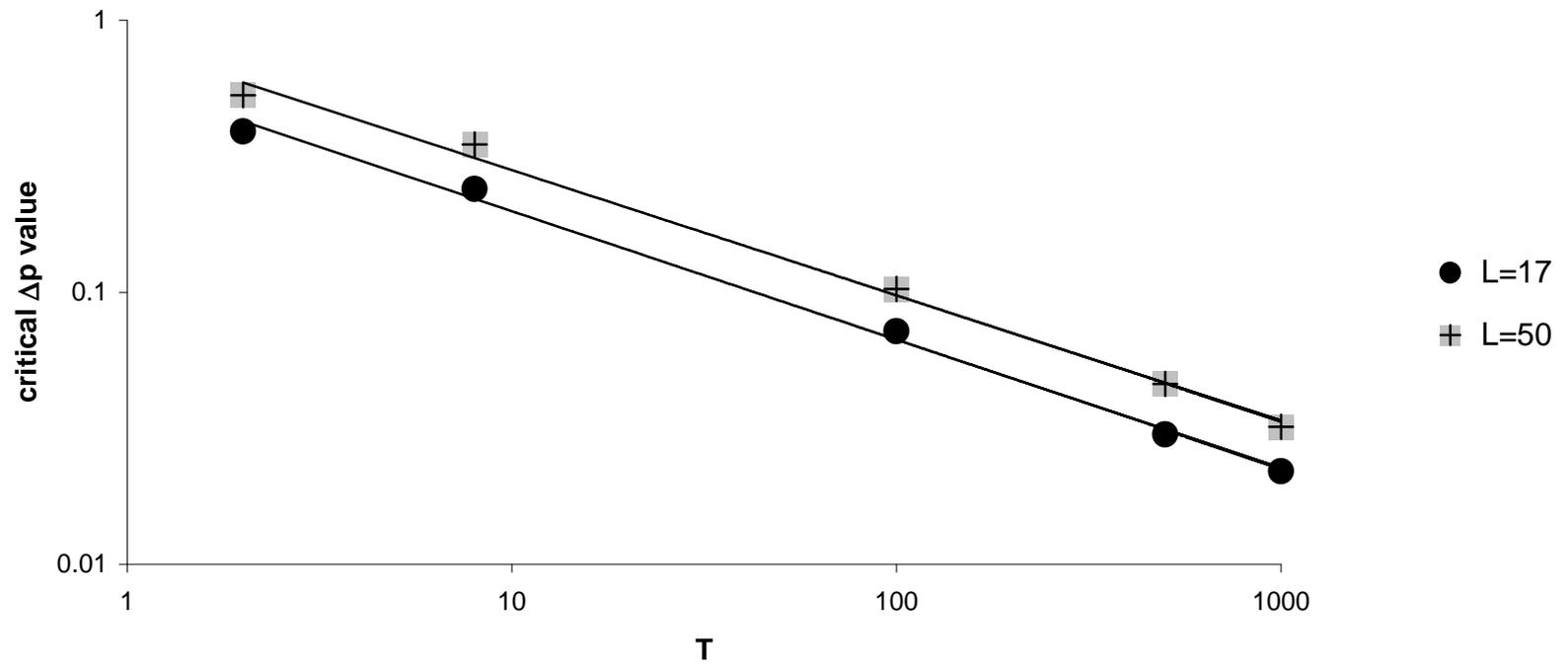
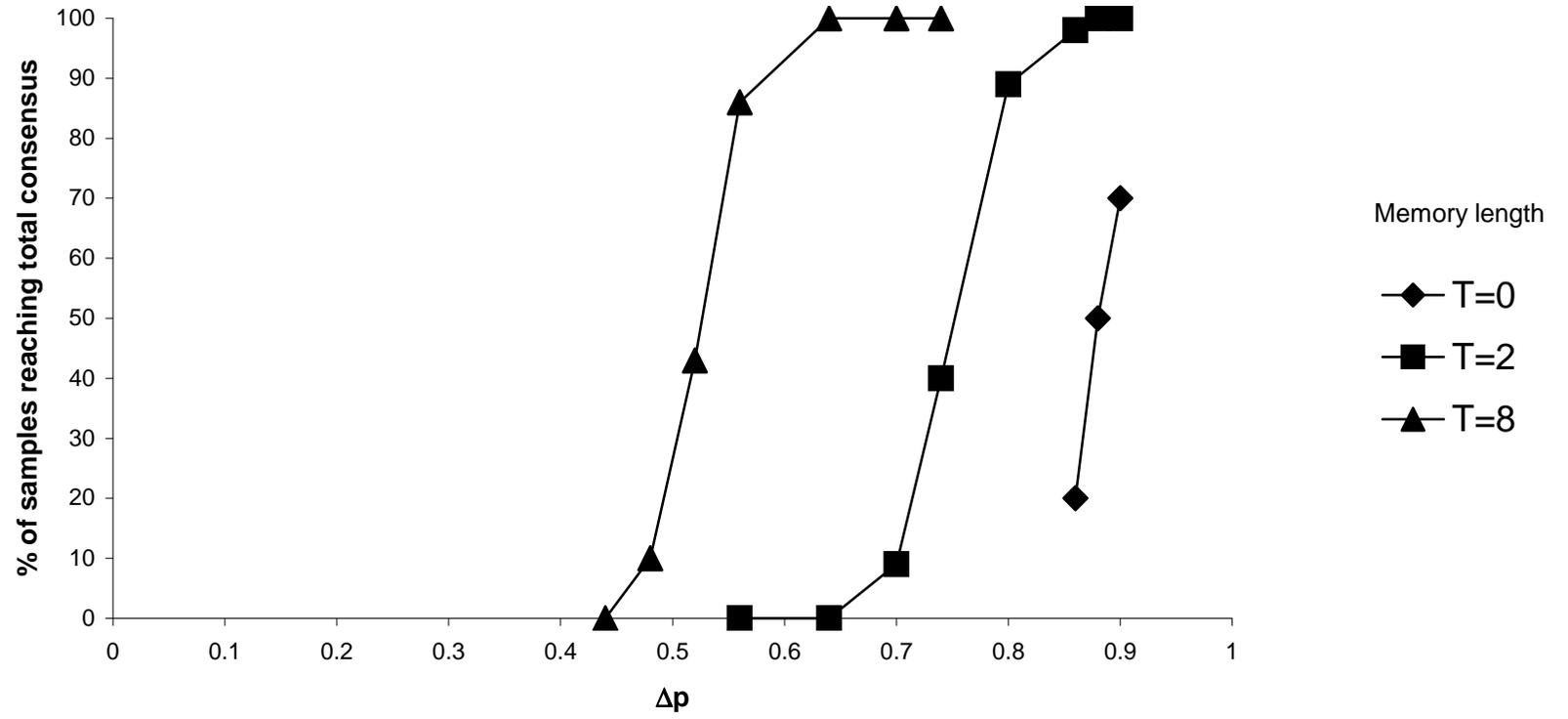
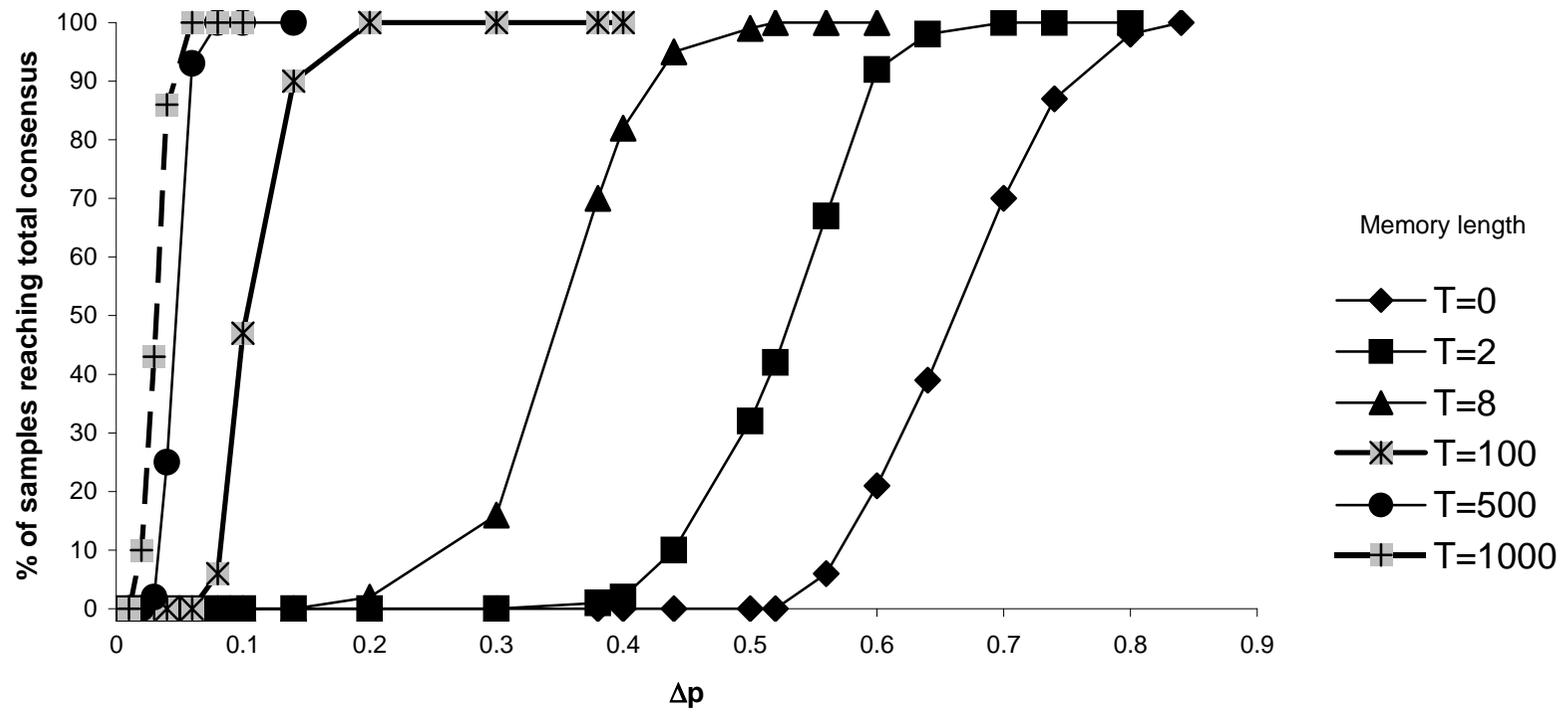
\end{document}